# Desorption of hydrocarbon chains by association with ionic and nonionic surfactants under flow as a mechanism for enhanced oil recovery

**Ketzasmin A. Terrón-Mejía[1], Roberto López-Rendón[1], Armando Gama Goicochea[2*]**

[1]Laboratorio de Bioingeniería Molecular a Multiescala, Facultad de Ciencias,
Universidad Autónoma del Estado de México,
Av. Instituto Literario 100, 50000, Toluca, Mexico

[2]División de Ingeniería Química y Bioquímica, Tecnológico de Estudios Superiores de Ecatepec,
Av. Tecnológico s/n, Ecatepec, Estado de México 55210, Mexico

The need to extract oil from wells where it is embedded on the surfaces of rocks has led to the development of new and improved enhanced oil recovery techniques. One of those is the injection of surfactants with water vapor, which promotes desorption of oil that can then be extracted using pumps, as the surfactants encapsulate the oil in foams. However, the mechanisms that lead to the optimal desorption of oil and the best type of surfactants to carry out desorption are not well known yet, which warrants the need to carry out basic research on this topic. In this work, we report non equilibrium dissipative particle dynamics simulations of model surfactants and oil molecules adsorbed on surfaces, with the purpose of studying the efficiency of the surfactants to desorb hydrocarbon chains, that are found adsorbed over flat surfaces. The model surfactants studied correspond to nonionic and cationic surfactants, and the hydrocarbon desorption is studied as a function of surfactant concentration under increasing Poiseuille flow. We obtain various hydrocarbon desorption isotherms for every model of surfactant proposed, under flow. Nonionic surfactants are found to be the most effective to desorb oil and the mechanisms that lead to this phenomenon are presented and discussed.

---

[*]Corresponding author. Electronic mail: agama@alumni.stanford.edu

**INTRODUCTION**

Recovery is at the heart of oil production from underground reservoirs. If the average worldwide recovery factor from hydrocarbon reservoirs can be increased beyond current limits, it will alleviate a few issues related to global energy supply [1]. The energy demand will be met by a global energy mix that is undergoing a transition from the current dominance of fossil fuels to a more balanced distribution of energy sources [2]. This challenge becomes an opportunity for technologies such as secondary and enhanced oil recovery (EOR) that can mitigate the balance between supply and demand [1]. EOR techniques can significantly extend global oil reserves once oil prices are high enough to make these techniques economically attractive [3]. Physically, the essence of the EOR methods lies in the reduction of interfacial tension between oil, surface and water thereby making it easy for oil to coalesce and flow out of the reservoir to production wells [4]. This is achieved using surfactants. The search for new and efficient models of surfactant molecules is challenging. For instance, Iglauer et al., proposed new surfactant classes for EOR using coreflood tests on Berea sandstones [5]; they documented how diverse theoretical and experimental studies can be used to propose new formulations of surfactants. Recent developments in surfactant EOR have greatly reduced the surfactant concentration required for effective oil recovery. Bera et al., reported that the adsorption of surfactants on rock surfaces can be reduced or altered by fixing the solution's pH for nonionic and ionic surfactants, which is an prominent issue regarding the economic feasibility for surfactant flooding [6]. Surfactant manufacturing is now delivering more advanced and safer EOR products at a lower cost than ever before. Early research on surfactants used in EOR focused on the injection of microemulsions into reservoirs. Microemulsions are also potential candidates in EOR, especially due to the ultra-low interfacial tension values attained between the contacting oil and water microphases that compose them [7].

Surfactants are commonly used in EOR processes for various purposes, including reduction of oil/water interfacial tension, wettability alteration, and foam generation [8]. The concurrent removal of surfactant and oil remains unsolved because the existing filtration membranes still suffer from low surfactant removal rate and due to surfactant-induced fouling [9]. For instance, in the surfactant – based remedial technologies, the cationic surfactants are more likely to adsorb onto the surface of negatively charged soil particles and aquifer materials, which inevitably increased the consumption of surfactants. Therefore, more cases that use anionic surfactants, instead of cationic surfactants, for soil washing or aquifer flushing were reported [10]. Cationic surfactants are, in general, more expensive than anionic surfactants because of the high-pressure hydrogenation reaction required during their synthesis, yet they are often of great commercial importance, such as in corrosion inhibition [11].For ionic surfactants, it is now widely accepted that there are significant changes in the free surfactant concentrations the total surfactant loading is increased above the critical micelle concentration. This is confirmed by experiments [12-14], simulations [15–17] and theory [18–20].The magnitude of the decrease in free surfactant concentration for nonionic surfactants is significantly smaller than that for ionic surfactants and still somewhat controversial [21]. Depending on the nature of the hydrophilic group, synthetic surfactants are classified in four types. The hydrophilic group is usually a sulphate group, a sulfonate group, or a carboxilate group (for anionic surfactants), a quaternary ammonium group (for cationic surfactants), polyoxyethylene, sucrose, or polypeptide (for nonionic surfactants). The most common hydrophobic parts of the synthetic surfactants are paraffins, olefins, alkylbenzenes, alkylphenols, and alcohols [22].

Despite recent advances in the development of new and more effective surfactants to improve the EOR process, the essential interactions between oil and surfactant molecules and the desorption mechanisms taking place are not yet fully understood. In this sense, computer simulations allow one to access parts of the system in detail, and the emerging quantitative results yield a link to the data from experimental approaches. Simulation models for science and technology have been developed from the early days of computers and have become—with the increasing performance of computers—a standard tool in physics, chemistry, applied sciences and engineering [23].

In this study, we undertake the challenge of the removal of hydrocarbons from surfaces through association with surfactants as a mechanism for EOR. Surfactant–polymer systems are investigated using dissipative particle dynamics (DPD) numerical simulations [24, 25] under constant Poiseuille flow. DPD is a mesoscopic method that employs soft repulsive potentials, enabling the sampling of particle – based systems with time steps that can be up to three orders of magnitude larger than those used in conventional molecular dynamics [26]. DPD simulations of surfactants have been carried out by several groups to understand their structural and thermodynamic properties. Recent reviews [27, 28] report studies dealing with the use of DPD for the prediction of the spatial arrangement of amphiphilic molecules in a solvent. Jury et al. [29] used DPD to simulate a dense solution of an amphiphilic species consisting of rigid surfactants in solution, finding a phase diagram that agreed with experimental data. Similarly, Nakamura [30] obtained the same phase diagram as Jury et al. [29] from his DPD simulation of amphiphilic molecules, and noted that the formation of the hexagonal phase takes longer to form than the lamellar phase. Nakamura and Tamura [31] investigated the hydrophilicity dependence of the phase structure more qualitatively by varying the interaction potential between hydrophilic molecules and water molecules, using DPD simulations. The dependence of the excess pressure and surface tension for binary mixtures of monomeric liquids were investigated, as functions of surfactant concentration at the interface [32]. It has been observed in the DPD simulations of Li et al. [33] on surfactant monolayers at the interface between oil and water that for decreasing the interfacial tension, it is better to have the same structure for the hydrophobic chains of surfactants and oil since the molecules at the interfacial layer could array compactly. Rekvig et al. [34] examined the effects of size and structure of surfactants on reduction of interfacial tension and the influence of branching of the hydrophobic tail. Also, Rekvig et al. [34] reported that branching has a positive effect on the efficiency of surfactants at the interface only if the head groups are sufficiently hydrophilic to prevent molecules from staggering. Amphiphilic monolayers on the interface between oil and water were simulated also by Rekvig et al. [35] using DPD, confirming that the thickness of the layer affects the rigidity more than the density of the layer. They also found that mixtures of short and long surfactants are more flexible than medium length surfactants of the same average chain length. Using the DPD model, Dong et al. [36] studied the orientation of sodium dodecyl sulfonate and sodium dodecylsulfate (SDS), and they found that strong hydrophilic head groups and the addition of salt results in the surfactants being stretched and ordered. Mesostructures and morphologies of mixed surfactant solutions containing cationic Gemini and anionic SDS surfactants were investigated by Wang et al. [37], who found a variety of mesostructures by controlling the conservative term in the DPD algorithm, which were consistent with experiments. From the DPD simulations of Yang et al. [38], different phase structures of an anionic surfactants mixed in water were observed, and the influence of concentration and temperature on the phase behavior of lamellar regions were studied. To explore the properties of polymer and surfactant systems, Yuan et al. [39] selected two cationic surfactants and a polymer in their DPD simulations, and in subsequent work [40], they carried out similar DPD simulations for the investigation of other surfactant–polymer mixtures.

Here we report extensive non–equilibrium DPD simulations of linear hydrocarbons (HC) adsorbed on flat surfaces as a model for oil embedded in rocks. To promote de desorption of the HC, a varying number of surfactant molecules dissolved in water is added to the system, as a mechanism for EOR. Two types of linear surfactants are used: a nonionic, and a cationic surfactant. In each case, five − bead and ten − bead surfactant chains are modeled. A crucial novel aspect of the work reported here is that all the particles within the fluid confined by the parallel, flat surfaces are subject to a constant, external force in the direction parallel to the plane of the surfaces, leading to stationary Poiseuille flow [41, 42]. As shall be discussed in what follows, increasing the value of the external flow promotes desorption of the hydrocarbon (HC) at smaller surfactant concentration, which means that flow is an important variable that should be considered when studying EOR mechanisms. Following experimental and numerical reports [43 - 48] we start by modeling oil as a mixture of mainly four hydrocarbons: butane, heptane, decane, and dodecane, with sixty molecules of each of those in the simulation box. Figure 1 shows the different molecules modeled in this work, as well as their coarse − grained DPD representation; Figs. 1 (A) - (D) display the HC molecules, which are mapped into four -, seven −, ten-, and twelve– bead linear DPD molecules, corresponding to $C_4H_{10}$, $C_7H_{16}$, $C_{10}H_{22}$, $C_{12}H_{26}$, respectively, which constitute our HC model. On the other hand, the solvent is modeled as a fluid made up of monomeric DPD particles. The interactions of these monomers with themselves and with the rest of the particles depends on the number of water molecules a DPD solvent particle is thought to encapsulate, in our case this (also called the "coarse − graining degree") is three water molecules, see Fig. 1 (E). The coarse − grained model for the nonionic surfactants are shown in Fig. 1 (F), with the one on the left side being the five – bead surfactant, and the one on the right side is the ten – bead one. The structures of the equivalent length for the ionic surfactants are shown in Fig. 1 (G). Further information about the model parameters and simulation details can be found in the **METHODS** section and full details are reported in the Supplementary Information (**SI**).

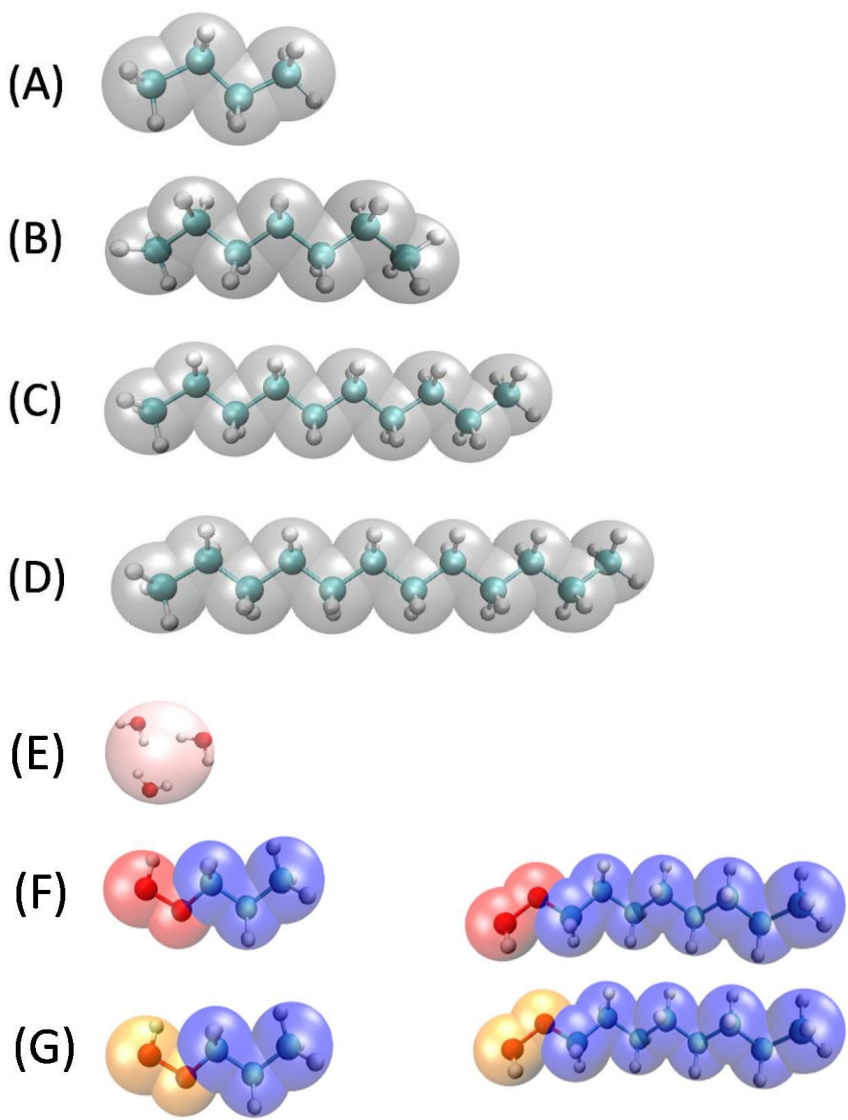

**Figure 1(Color online) Schematic representation of the coarse-grained models adopted in this work.** (A) Butane model; it is mapped into a four – bead linear molecule of identical DPD beads. (B) Model for the heptane molecule. (C) Decane molecule model. (D) Dodecane molecule model. (E) Water DPD model. (F) Models for non-ionic surfactants. The five – bead chain is shown on the left, and on the right is the ten – bead surfactant. The red beads (left end of the molecules) represent the hydrophilic head and blue beads correspond to the hydrophobic tail. (G) DPD models for ionic surfactants, which are of the same length as their nonionic counterparts. The yellow beads (left end of the molecules) represent the hydrophilic head and blue beads correspond to the hydrophobic tail. These figures were prepared with VMD [49].

**RESULTS AND DISCUSSION**

To obtain desorption isotherms we start by fixing the number of HC molecules in the simulation cell; those molecules adsorb on the surfaces because they are dissolved in an aqueous solvent and they are subject to hydrophobic attraction for the walls. After all HC molecules have been adsorbed we add surfactants in increasing concentration to each simulation cell, until desorption of the HC molecules is observed. In Fig. 2 we show snapshots of several cases used to obtain desorption isotherms when long non − ionic surfactants are used, from the low surfactant concentration case (Fig. 2A), to the almost complete desorption of HC molecules (Fig. 2D). The volume is the same for all simulations, and they are all under the influence of external Poiseuille flow, unless stated otherwise.

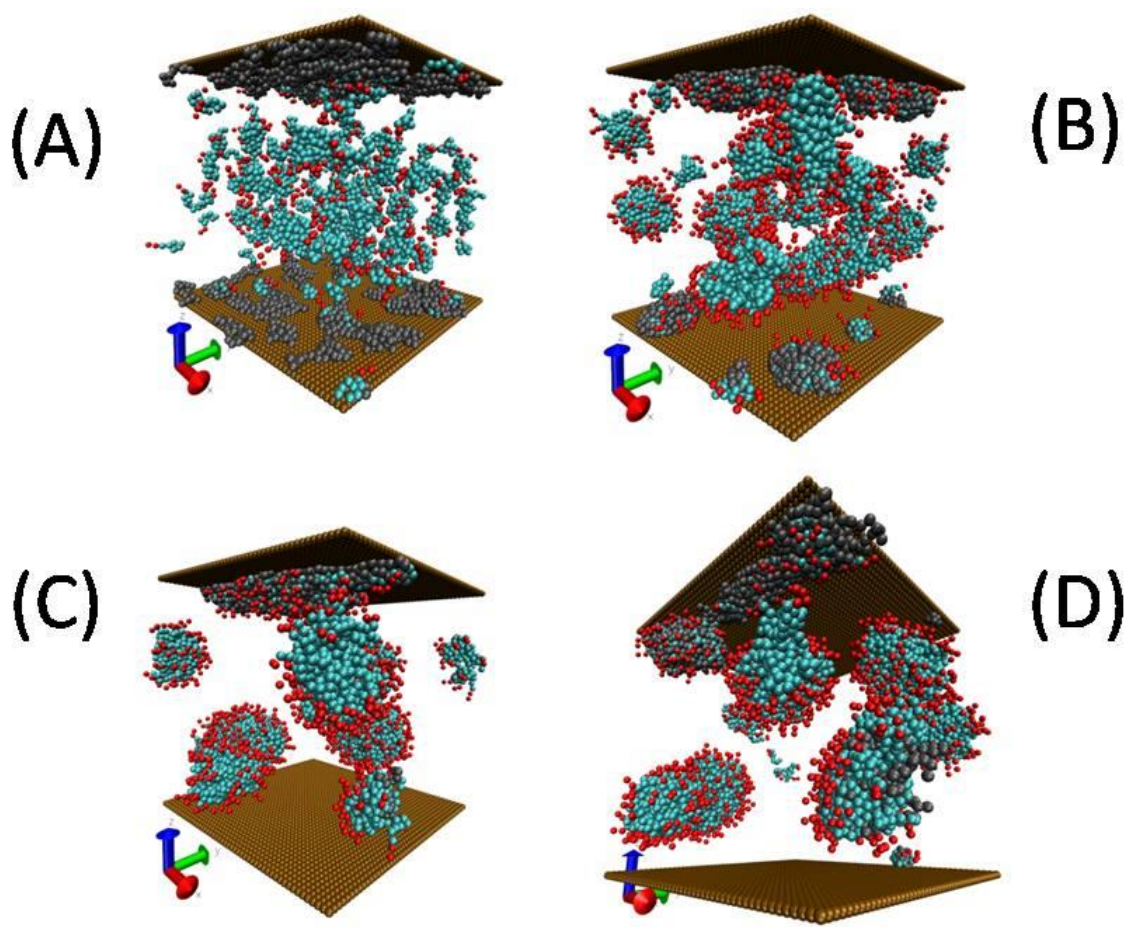

**Figure 2 (Color online) Snapshots of the encapsulation of HC through association with short chain non-ionic surfactants.** These snapshots were taken from final configurations of the simulation at increasing surfactant concentration. Gray molecules represent the mixture of HC chains, while cyan beads correspond to the hydrophobic tail of the surfactants, and read beads are the hydrophilic heads. The surfactant – driven desorption process is seen in (C) and (D).

At low concentration, all surfactant chains associate with the adsorbed HC molecules, see Fig. 2B. However, as the concentration is increased there appears the incipient formation of surfactant micelles, while the rest of the surfactant molecules continue associating with the adsorbed HC molecules, see Fig. 2C. This competition between adsorption through association with polymers, and self – association in micelles has been observed in experiments on competitive adsorption [50-52] as well as in equilibrium simulations of complex fluids with polymers and surfactants [53]. Increasing the surfactant concentration leads eventually to the almost complete desorption of the HC chains (see Fig. 2D), because most of them are surrounded by the hydrophobic surfactant tail, creating complex associated particles whose outer components are the hydrophilic heads of the surfactants, which have been attracted to the aqueous fluid and desorb from the surfaces.

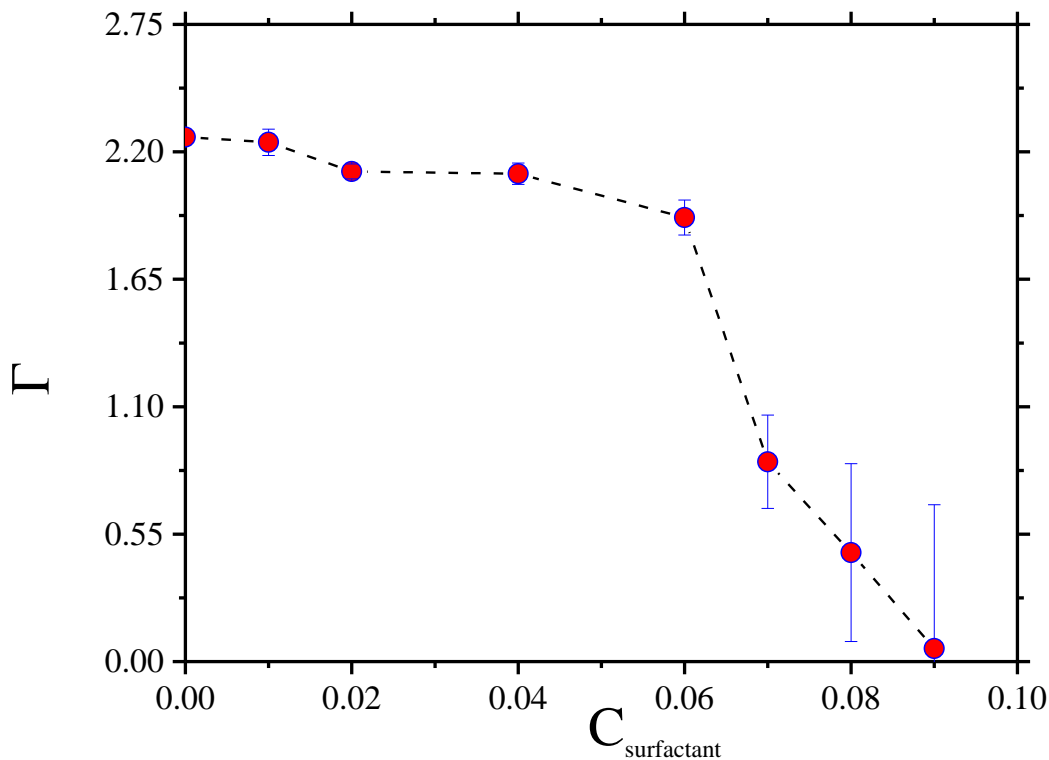

**Figure 3 (Color online) Desorption isotherm ($\Gamma$) for a mixture of four types of HC (butane, heptane, decane and dodecane, with molecules of each in the cell) by means of the addition of long nonionic surfactants, see right image in Fig. 1F, as a function of surfactant concentration ($C_{surfactant}$).** The line is only a guide for the eye; all quantities are reported in reduced DPD units.

Following the procedure described above one can construct HC desorption curves, which are shown in Fig. 3, where the $x-$ axis shows the surfactant concentration in each simulation cell, and the $y-$ axis represents the total number of HC molecules adsorbed per unit area ($\Gamma$) . Each data point represents a simulation cell with a given surfactant concentration. Only long, nonionic surfactants were used to obtain this desorption isotherm, see the image on the right side in Fig.

1F. There appears a plateau in the desorption isotherm as the surfactant concentration is increased, because at relatively low concentration the surfactants associate with the HC molecules but there are not enough surfactants to desorb the HC molecules, see Figs. 2(A) and 2(B). On further increase in the surfactant concentration, the HC molecules become encapsulated by the surfactants and desorption from the surfaces occur.

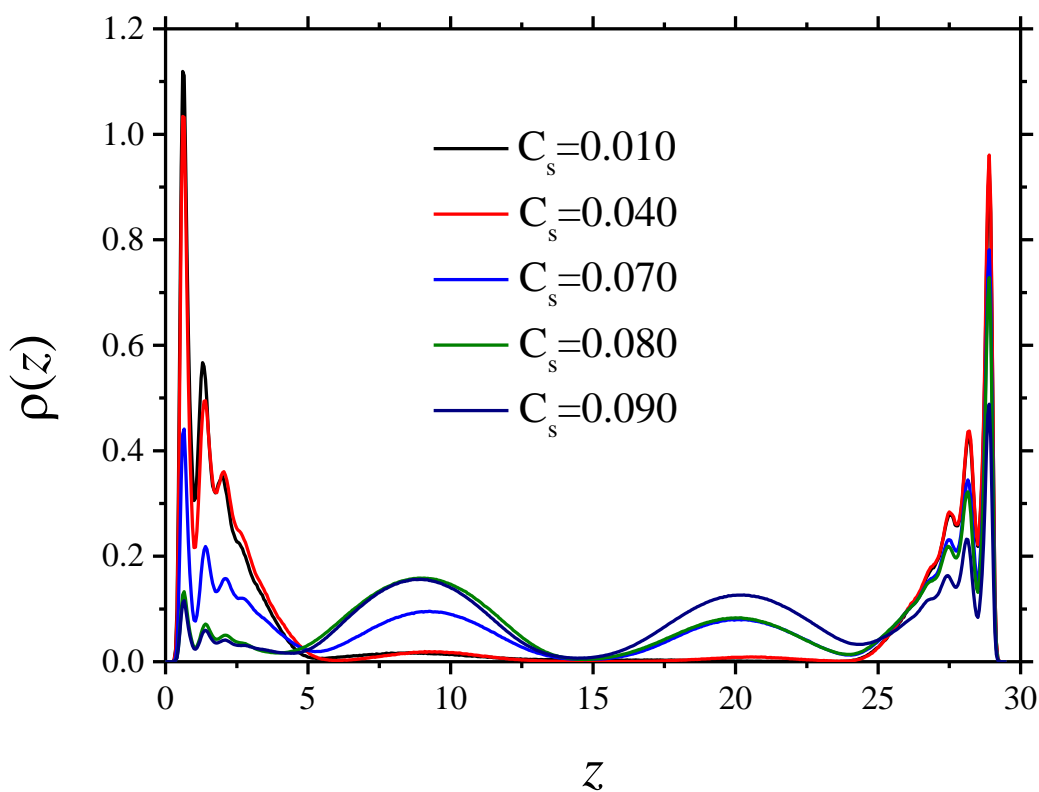

**Figure 4 (Color online) Hydrocarbon molecules concentration profiles for increased long, nonionic surfactant concentration.** At low surfactant concentration ($C_s$) no desorption is observed, but as $C_s$ is increased the HC begin to desorb, see the blue line. All quantities are reported in reduced DPD units.

Figure 4 shows the concentration profiles of the HC molecules for three different long, nonionic surfactant concentrations, $C_s$. At relatively low values of $C_s$ there is virtually no HC desorption, and they form layers near the surfaces, as indicated by the maxima in Fig. 4. Increasing the surfactant concentration leads to desorption of HC molecules, which is signaled in Fig. 4 by the maxima of the blue line ($C_s$=0.070) at $z = 8$ and $z = 20$. To trace further the desorption mechanism and reduce variables we focus in the rest of this work on systems that contain only heptane molecules, and study instead the influence of the surfactant chain length and its electrostatic nature (ionic vs. nonionic). Figure 5 (A) shows desorption isotherms when nonionic surfactants are used, with the circles (red dashed line) corresponding to long chains. The squares (blue) in Fig. 5 (A) represent heptane desorption when short nonionic surfactants are used. In Fig. 5 (B) we see the equivalent heptane desorption isotherms, for ionic surfactants. The solid squares in Fig. 5 (B) make up the isotherm obtained with short ionic surfactants, and the solid circles are the desorption curve for long – chain ionic surfactants.

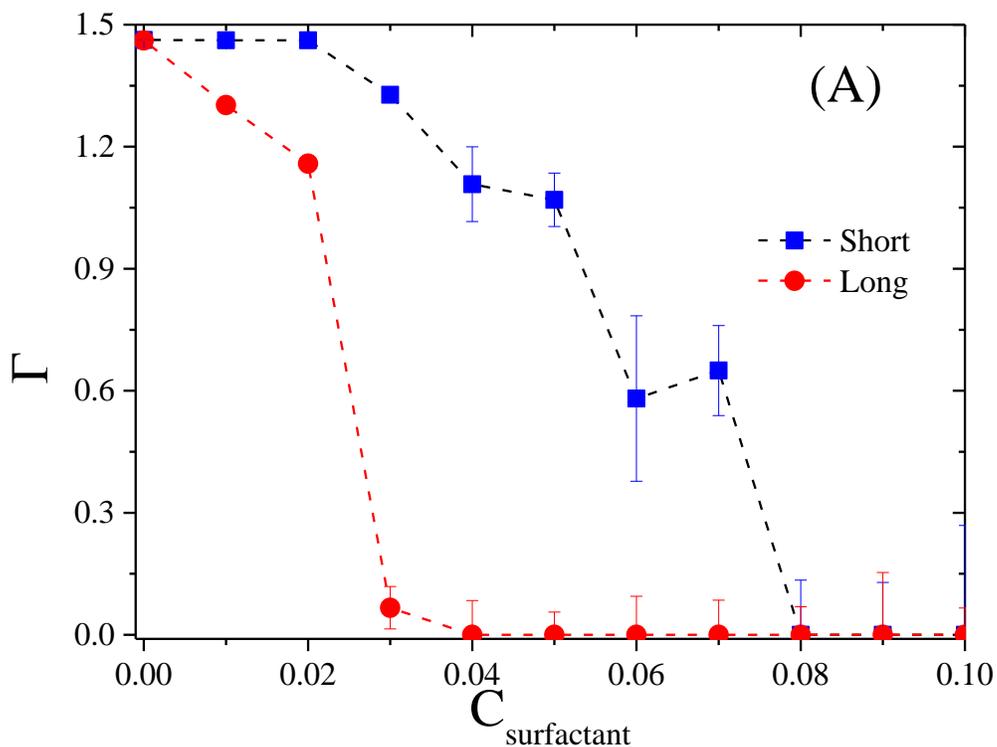

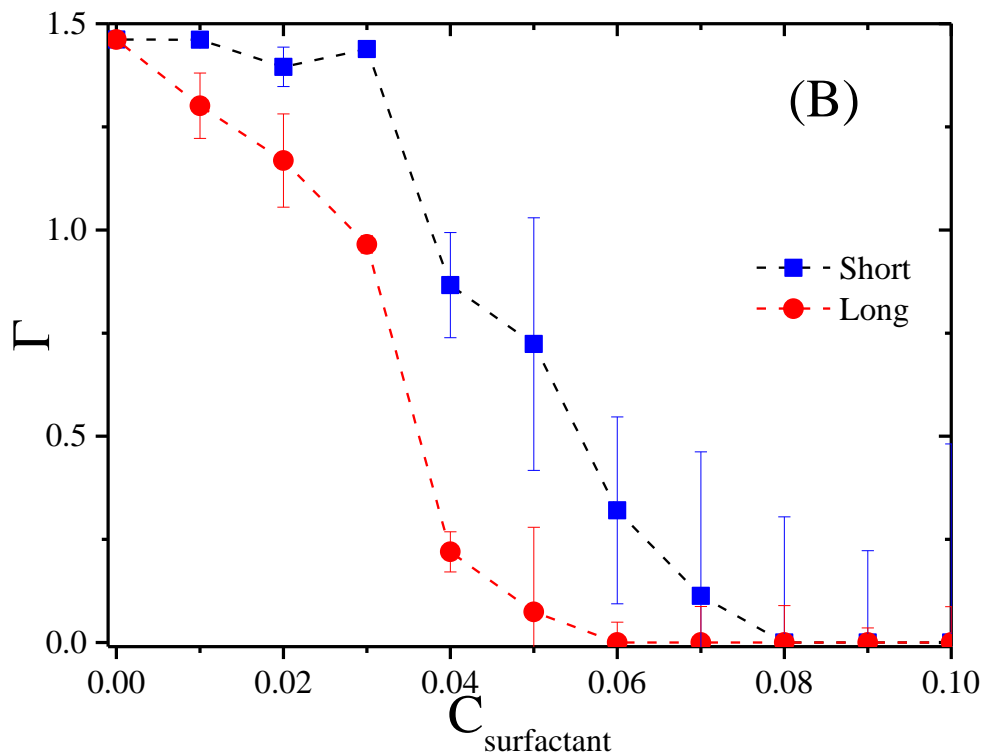

**Figure 5 (Color online) Desorption isotherms ($\Gamma$) for heptane with nonionic and ionic surfactants as functions of surfactant concentration ($C_{surfactant}$).** (A) The red line (circles) and blue line (squares) correspond to long and short nonionic surfactant chains, respectively. (B) Desorption isotherms with long (circles) and short (squares) ionic surfactants. Lines are only guides for the eye; all quantities are reported in reduced DPD units.

Comparison of long and short surfactants is carried out at the same mass concentration, rather than at the same molar concentration. Therefore, the number of beads that make up surfactant molecules is the same at a given value of $C_{surfactant}$, regardless of whether they belong to long or short surfactant chains. Since the volume of the simulation cells and the number of heptane molecules are the same in all simulations, all curves shown in Fig. 5 start from the same point. However, as the surfactant concentration is increased one finds that the long – chain, nonionic surfactants are the most effective in desorbing the HC molecules from the surfaces, see the solid circles in Fig. 5 (A), since they require the smallest concentration to produce complete desorption of the HC. By contrast, long – chain ionic surfactants require an increase of about twenty percent in their concentration to accomplish complete desorption of the HC. For both surfactant types, ionic and nonionic, Fig. 5 shows that short chains require increased concentration to promote complete desorption of heptane when compared with their long – chain counterparts, which is expected since more chains would be needed to compensate the hydrophobic interaction of the short surfactant tails with the heptane beads. A qualitative difference between the heptane desorption mechanisms with nonionic and ionic surfactants is that in the latter there is competition between the hydrophobic and electrostatic interactions between the charged groups, which leads to the requirement of more ionic surfactants to accomplish full desorption. While for nonionic surfactants the desorption mechanism involves only the competition between surface – HC – surfactant and surfactant – HC –surfactant interactions that lead to adsorption and micelle formation [53], for ionic surfactants there appears also the electrostatic attraction between the charged groups in the surfactants and their counter ions. Additionally, there is electrostatic repulsion between the ionic surfactants that associate with the HC molecules, and this is turn leads to higher surfactant concentration before complete desorption of HC takes place, as shown in Fig. 5 (B). These results follow trends found previously in simulations of polyelectrolyte adsorption [54].

The density profiles of heptane molecules along the direction perpendicular to the surfaces shed light on the desorption mechanisms described previously, as they show how heptane molecules arrange themselves according to the composition of the fluid that surrounds them, as seen in Fig. 6.

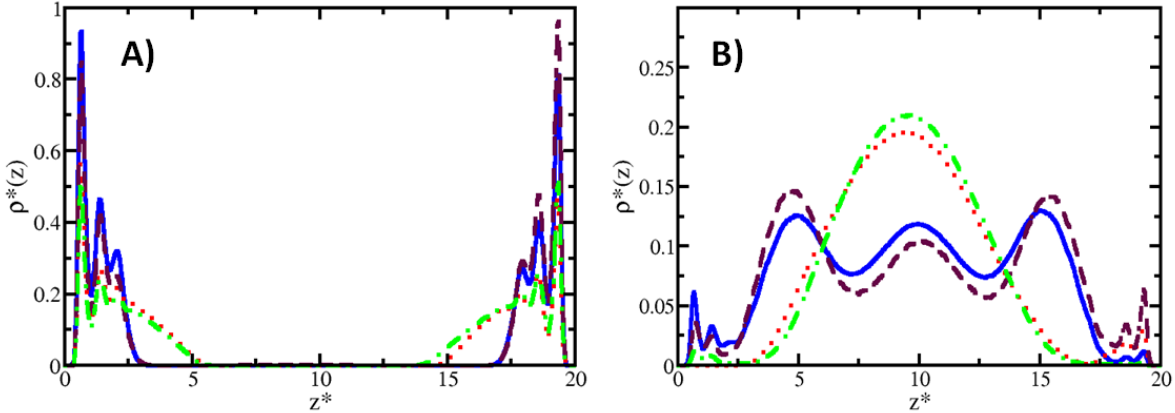

**Figure 6 (Color online) Density profiles of heptane before and after the desorption process by means of nonionic and ionic surfactants.** The red and blue lines correspond to the cases when long and short nonionic surfactants were used, respectively; the green and burgundy lines correspond to the cases when long and short ionic surfactants, respectively. Note that the density profiles of the surfactants are not shown, to improve the clarity of the figures. **A)** Density profiles *before* the desorption process of HC occurs. These profiles were obtained for a nonionic surfactant concentration equal to 0.02 (red and blue lines), and equal to 0.03 for the case of ionic surfactants (green and burgundy lines). These are the threshold values of the concentration at which desorption begins to occur, see also Fig. 5. **B)** Density profiles *after* desorption of HC. These profiles were obtained for a concentration of 0.04 of long nonionic surfactants, and of 0.08 for the case of short non-ionic surfactants (see Fig. 5 (A)). The profiles for ionic surfactants correspond to a concentration of 0.06 for long chains and 0.08 for the case of short chains (see Fig. 5 (B)). All quantities are reported in reduced DPD units.

Figure 6A shows the heptane density profiles at surfactant concentration just below desorption, for both ionic and non – ionic surfactants, of both lengths. The peaks correspond to the structuring of the HC in layers near the surfaces; notice the HC is adsorbed symmetrically on both walls. Another salient feature is that for the ten – bead surfactants, the HC extends more into the centre of the pore (green and red lines in Fig. 6A, for ionic and non – ionic surfactants, respectively) than for the five – bead surfactants (burgundy and blue lines in Fig. 6A, for ionic and non – ionic surfactants, respectively). Therefore, long surfactants are more effective in desorbing HC molecules than shorter ones, see also Fig. 5. In Fig. 6B we present the HC density profiles at surfactant concentrations where desorption of heptane has occurred, for all four cases studied here. The green and red lines show maxima in the centre of the pore and they are nearly zero close to the surfaces, indicating that heptane has been completely desorbed by long ionic surfactants (green) and long non – ionic surfactants (red). The HC density profiles using short ionic (burgundy line) and non – ionic (blue line) surfactants to desorb are shown in Fig. 6B. Notice how, in these cases, there is a more uniform distribution of heptane along the pore defined by the parallel walls with three maxima appearing, instead of the single maximum found when long surfactants are used. The profiles shown in Fig. 6B are complex structures made up of heptane molecules, surrounded by the hydrophobic groups of the surfactants; the outer layers of these composite particles are the hydrophilic heads of the surfactants, interacting with the solvent. To illustrate the formation of these structures we show in Fig. 7 snapshots of the final configurations, namely those where the HC molecules are completely desorbed from the surfaces, focusing on the case of nonionic surfactants, for clarity.

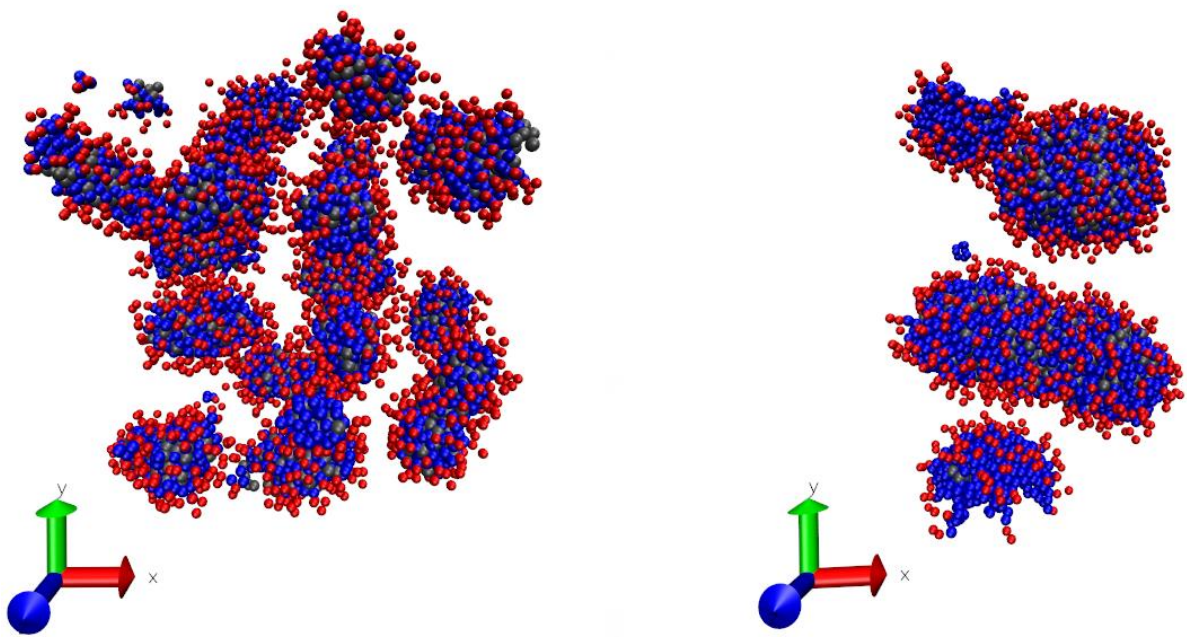

**Figure 7 (Color online) Snapshots of the final configuration for the HC desorption process using nonionic surfactants.** On the left is the short surfactant case, while the right image corresponds to long chain surfactants. Note how the HC molecules are completely covered by the surfactant molecules. The solvent and the surfaces are omitted for clarity.

The snapshot on the left in Fig. 7 corresponds to desorption of HC with short – chain, nonionic surfactants, while the one on the right is the long – chain, nonionic surfactant case. Short surfactants promote the formation of smaller composite particles (surfactant – HC) that lead to HC desorption, maximizing their translational entropy at the cost of increased surfactant concentration. Long surfactants, on the other hand, can encapsulate more HC molecules with fewer surfactant molecules, therefore a smaller concentration is required to desorb completely the HC, which leads to the desorption isotherm shown in red circles in Fig. 5 (A).

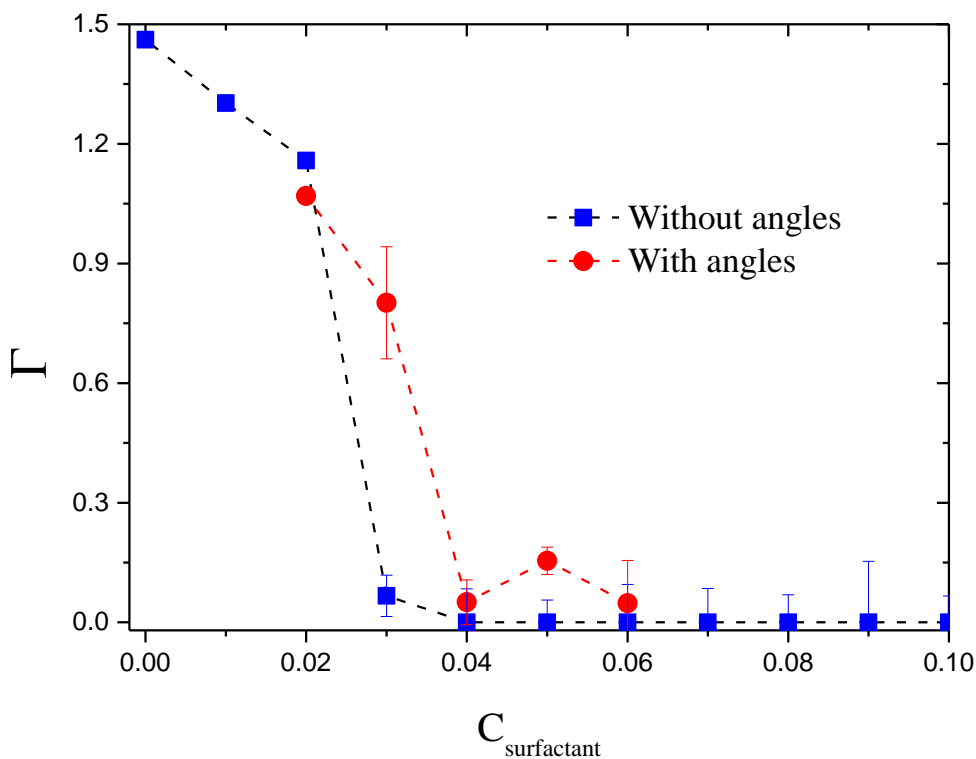

**Figure 8 (Color online) Heptane desorption isotherms with, and without bond angle interactions using long, nonionic surfactants.** Solid blue squares are those for the case when the monomers in the heptane molecule are joined by freely rotating springs (taken from Fig. 5 (A)), while the red circles represent the case when there is a three-body harmonic potential between bonds along the heptane molecule.

Coarse – grained simulations such as those reported here group several atoms into a single particle whose size can be larger than the persistence length of the polymer chain, as is the case here [26, 28, 34]. This means that there is no need to include bond angle an dihedral interactions [46] at the relatively large coarse – graining degree in DPD we use in this work. To show this is the case we have performed additional simulations in which three – body interactions between neighboring bonds along the heptane chain were included (all details are provided in the Supplementary Information). Figure 8 shows the comparison of desorption isotherms obtained when three – body interactions are included (circles in Fig. 8) and when they are not (squares in Fig. 8); only long, nonionic surfactants were used in both cases. Although there are minor differences, such as in the desorption onset and some fluctuation, the qualitative trends are the same. Since there appears to be no fundamental influence of bond angle interactions in the HC desorption mechanism, and their inclusion at our coarse – graining degree is superfluous, they were not included in any other case.

To investigate the desorption mechanisms when ionic and nonionic interactions compete simultaneously we performed an additional series of simulations where the surfactants added to the system were a 50:50 mixture of long, ionic and nonionic surfactants under the same conditions as in the previous simulations. By carrying out simulations at increasing concentration of the mixture of surfactants, we calculated the heptane desorption isotherm, and compared it with those obtained for the pure nonionic and pure ionic surfactant cases. The results can be found in Fig. 9.

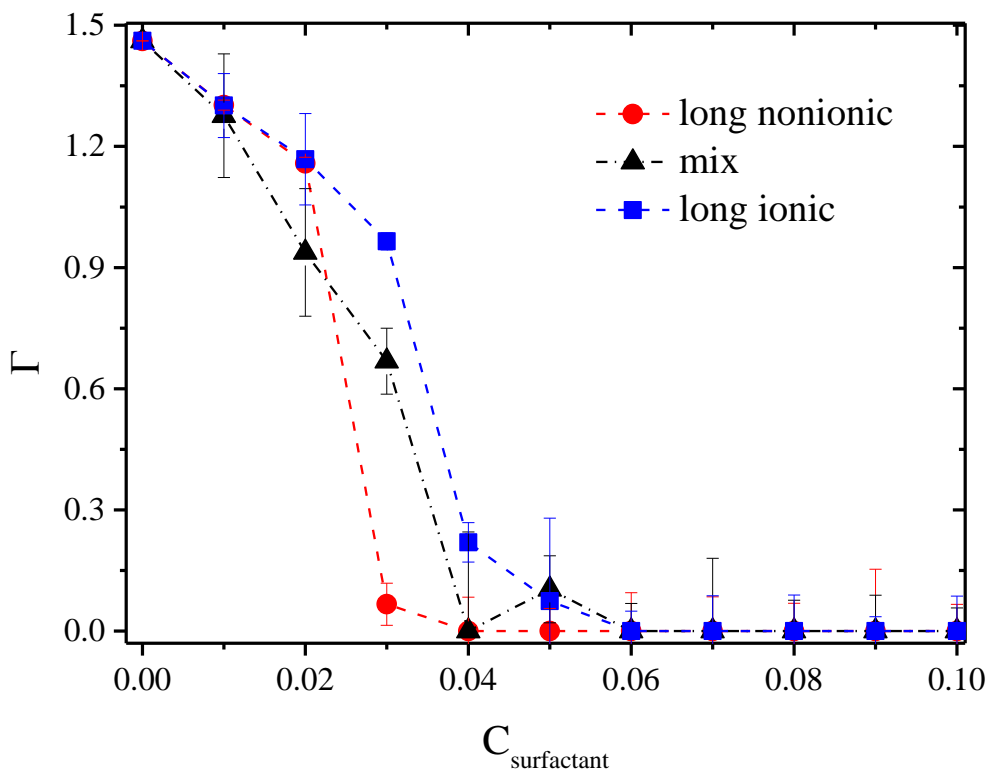

**Figure 9 (Color online) Desorption isotherms (Γ) of heptane using a 50:50 mixture of nonionic and ionic surfactants**. The red circles (nonionic surfactants) and blue squares (ionic surfactants) are the same isotherms presented in Fig. 5A and 5B, respectively, for long surfactants and are included here to emphasize the effect of the surfactant mixture. The black triangles correspond to a 50:50 mixture of ionic and non-ionic surfactants.

Figure 9 shows that adding a 50:50 mixture of long ionic and nonionic surfactants to the fluid made up of heptane adsorbed on surfaces and the solvent, does reduce the concentration of long ionic surfactants needed to desorb the HC completely by about twenty percent with respect with the pure ionic surfactant case, but it also creates fluctuations in the concentration of nonionic surfactants needed for complete desorption. This means that the addition of electrically charged molecules to the neutral system reduces the efficiency of the long nonionic surfactants to encapsulate the HC molecules due to the electrostatic repulsion between the ionic surfactant heads. Therefore, it appears to be unnecessary to use nonionic/ionic surfactant mixtures to desorb more effectively HC molecules.

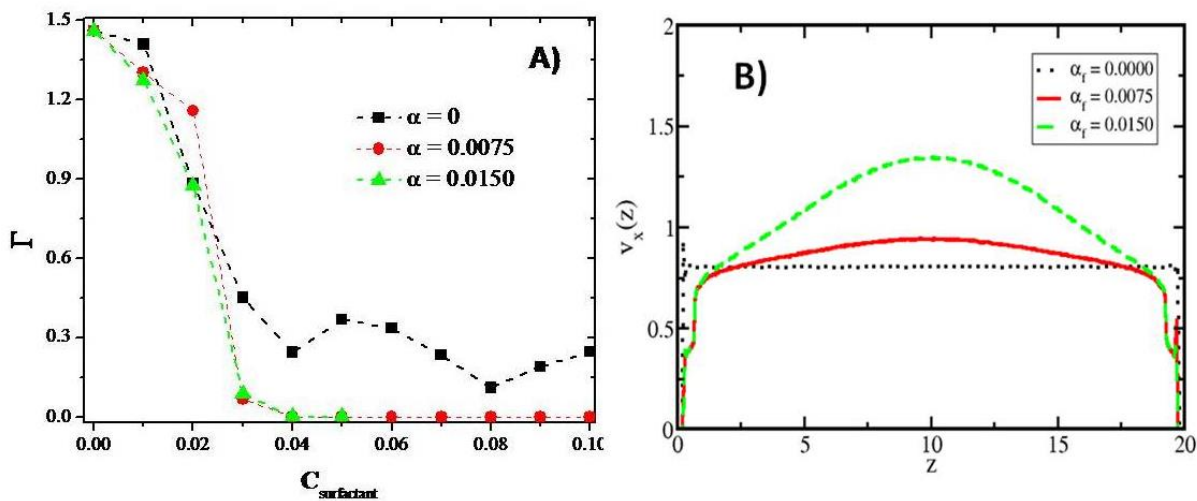

**Figure 10 (Color online) Influence of the flow rate in the desorption process of HC. A)** Desorption isotherms (Γ) of HC by long nonionic surfactants under three different flow conditions $(\alpha_f)$. The red line is the same isotherm as that presented in Fig. 5A, and is included here for comparison. **B)** Velocity profiles of desorption of HC by long chain nonionic surfactants under three flow conditions $(\alpha_f)$. For **A)** and **B)** the black (squares in Fig. 10 A), red (circles in Fig. 10 A) and green (triangles in Fig. 10 A) lines correspond to flow conditions equal to $\alpha_f = 0.0000, 0.0075$ and $0.0150$, respectively. See Eq. S11 of the **SI** for details about flow conditions$(\alpha_f)$. All quantities are reported in reduced DPD units.

Lastly, and considering that EOR is a procedure that is performed under flow, we study the influence of the strength of the external flow imposed on the fluid. All results presented so far correspond to a constant value of the external force that creates Poiseuille flow along the $x-$ direction, in particular, $F_{ext}(x) = \alpha_f$ with $\alpha_f = 0.0075$ in reduced DPD units. In Fig. 10A, we show the comparison between three Poiseuille flow conditions focusing only on the optimal case, i.e. for long nonionic surfactants, where the black line (squares) corresponds to zero flow, the red line (circles) is the same curve as that shown in red in Fig. 5A (with $\alpha_f = 0.0075$), and the green line (triangles) is the isotherm obtained at $\alpha_f = 0.0150$. Under equilibrium (zero flow) conditions, even the optimal surfactant concentration is incapable to desorb all HC molecules, but as soon as an external force is applied nonionic surfactants become a useful additive as a means of EOR. The onset of desorption in surfactant concentration is reduced when the external flow is increased, as the green line in Fig. 10A shows; this is because many HC molecules have been successfully encapsulated by the surfactants, but some remain adsorbed because they are not completely covered yet by the surfactants. Those weakly adsorbed HC molecules, see for example, Fig. 2D, can more easily be desorbed if the magnitude of the external flow is increased, which is indeed what one sees in the green line (triangles) in Fig. 10A. However, what could be called the "critical surfactant desorption concentration" (cdc) does not change ($C_{surf} = 0.04$ in Fig. 10A and Fig. 9), which implies that the cdc is set by the interactions among the constituents of the system and their molecular structure, but not necessarily by the flow conditions. Finally, Fig. 10B shows the profiles of the $x-$ component of the velocity of the particles in the fluid along the direction perpendicular to the surfaces. When the system is in equilibrium (no flow), the velocity profile along the pore is constant, see black dotted line in Fig. 10B, as expected. If there is external flow, the velocity profile adopts a parabolic shape, see solid and dashed lines in Fig. 10B, as is well known to occur under Poiseuille flow [55]. The stronger the flow, the more parabolic the velocity profile becomes, in agreement with simulations performed by others, see [56] and reference therein.

In conclusion, we have shown that the association of nonionic surfactants with heptane is a useful mechanism to promote the desorption of hydrocarbon molecules from surfaces where they are embedded, as occurs in many oil reservoirs, which works best when the fluid is subjected to constant external flow. Ionic surfactants were shown to be not as effective as nonionic ones in desorbing heptane, mainly because the electrostatic repulsion between the surfactants' heads leads to larger concentrations being required to desorb the hydrocarbon molecules. Hence, desorption is promoted by the reduction of the interfacial tension between hydrocarbons and the surface driven by surfactants, and enhanced by the external flow.

**METHODS**

The DPD model was used to carry out numerical simulations in the canonical ensemble (constant density and temperature), with the global density set equal to three, as is usually done. The electrostatic interactions were included using Ewald sums with charge distributions centered within the DPD beads. The surfactant and hydrocarbon molecules were constructed as chains of beads joined with freely rotating harmonic springs, and in one case a three – body angle harmonic potential between bonds was also included. The surfaces confining the fluid were defined as effective short – range forces. All computational details, model parameters, components of every system and full details of the work can be found in the SI.

**ACKNOWLEDGMENTS**

This work was supported by SIyEA-UAEM (Projects 3585/2014/CIA and 3831/2014/CIA). KATM acknowledges CONACYT for supporting his PhD studies. All simulations reported in this work were performed at the Supercomputer OLINKA located at the Laboratory of Molecular Bioengineering at Multiscale (LMBM), at the Universidad Autónoma del Estado de México. The authors are grateful to *Red Temática de Venómica Computacional y Bioingeniería Molecular a Multiescala* .AGG is indebted to J. A. Arcos for educational conversations.

## AUTHOR CONTRIBUTIONS

AGG conceived and designed the project; KATM wrote the code and performed the simulations, under the supervision of RLR and AGG. Funding for the project was secured by RLR. All authors contributed to the writing of the manuscript.

## ADDITIONAL INFORMATION

The authors declare no competing financial interests. Supplementary information accompanies this paper at www.nature.com. Reprints and permission information is available online at http://npg.nature.com/reprintsandpermissions/. Correspondence and requests for materials should be addressed to AGG.